# Experimental Software Schedulability Estimation For Varied Processor Frequencies

**Sampsa Fabritius[1]   Raimondas Lencevicius[2]   Edu Metz[2]   Alexander Ran[2]**

[1]*Nokia Mobile Phones, Sinitaival 5, FIN-33720 Tampere, Finland*
[2]*Nokia Research Center, 5 Wayside Road, Burlington, MA 01803, USA*
Sampsa.Fabritius@nokia.com    Raimondas.Lencevicius@nokia.com
Edu.Metz@nokia.com    Alexander.Ran@nokia.com

**Abstract**. *This paper describes a new approach to experimentally estimate the application schedulability for various processor frequencies. We use additional workload generated by an artificial high priority routine to simulate the frequency decrease of a processor. Then we estimate the schedulability of applications at different frequencies. The results of such estimation can be used to determine the frequencies and control algorithms of dynamic voltage scaling/dynamic frequency scaling (DVS/DFS) implementations. The paper presents a general problem description, the proposed schedulability estimation method, its analysis and evaluation.*

## 1   INTRODUCTION

Energy is an increasingly important resource for mobile devices necessitating the analysis and optimization of power consumption in all components of the mobile device. One possible way to achieve energy optimization in device's processor is DVS/DFS ([3][4][5][8][10]–[16])—dynamic voltage scaling/dynamic frequency scaling. Such scaling adjusts the operating voltage and the clock frequency of a system dynamically to fit the actual need. The processor load typically varies a lot and the processor does not need to run at the maximum frequency and voltage all the time. The tasks that require the most performance or the shortest latencies are executed at the maximum voltage and frequency. The tasks that do not have high requirements can be executed at lower voltage and frequency. The power consumption can be approximated by the equation $P = cCV^2F$ and the energy consumption by the equation $E = Pt$. From there $E = cCV^2Ft$. (c is constant factor, C is constant capacitance, V is voltage, F - frequency, t - time.) The reduction of voltage requires a proportional reduction in frequency. Assuming that the task would take proportionally longer to execute in lower frequency, the energy ratio in high voltage-high frequency execution and low voltage-low frequency execution is proportional to the square of the voltage ratio in these two executions ($V_{high}/V_{low}$)$^2$.

To summarize, it is possible to save energy by reducing the voltage and the frequency of a processor if the lower clock rate is sufficient for the applications, i.e. if there is enough time to execute them at the lower rate. This means that applications' executions have a property—*schedulability*—that indicates if application can tolerate DVS/DFS and what frequency they can tolerate.

There has been a lot of research in applying DVS/DFS to mobile systems and their software. It is important to indicate two directions that are not goals of this paper: First, we are not proposing a new approach to energy saving. We are not arguing for a new way to decrease the power consumption, we are using a well-known and researched DVS/DFS energy saving approach. Second, we are not proposing a new scheduling or DVS/DFS control algorithm. There have been a number of such algorithms proposed ([3][4][5][8][10]–[16]). We assume that one of them is used in the DVS/DFSed system. The goal of this work is orthogonal to the two directions above. It is to find out at what frequencies a system or an application is *schedulable* and consequently able to work under DVS/DFS. The paper explores this important quality, which enables power savings under DVS/DFS.

First, we present the general schedulability problem and possible solutions. Section 3 proposes a new schedulability estimation approach. Section 4 analyzes the advantages and drawbacks of the proposed method. We finish with the related work, the future work, and conclusions.

## 2   PROBLEM DEFINITION

This section formalizes the application schedulability problem, i.e. the problem of determining if it is possible to execute an application at a certain frequency.

Consider an event-driven system with interrupts and tasks of different priorities. Events occur in this system, some of them stochastically and some deterministically. Each event has the processing time and a deadline. The processing time indicates how long in seconds or how many processor cycles it takes to process the event. The deadline indicates a time limit by which an event has to be processed or else its processing constraint is violated. The problem then can be stated as follows: can the system be scheduled to obey the event deadlines at a certain processor frequency f? Changing processor frequency does not change the event processing times in processor cycles or the event deadlines in seconds. However, such frequency change increases event processing times in seconds.

A system is schedulable at frequency f if all its applications, possibly executed concurrently, are schedulable at frequency f. An application is schedulable at frequency f if all its executions are schedulable at frequency f. From here on we will discuss only schedulability of single applications and allowed application combinations, since a system is schedulable if all these cases are schedulable.

Without considering pathological cases, applications are schedulable at some frequency. The issue then is to find a minimum frequency at which they are schedulable. If an application is schedulable at a frequency lower than the maximum processor frequency, this application can be executed under DVS/DFS. In practice, systems such as mobile devices may not be schedulable at the lowest frequency derived from some theoretical model. For example, the Palm™ [9] operating system does not contain an advanced real-time scheduler that would be able to schedule an application at very low frequency achievable with an ideal scheduler. This means that analytical problem solutions should incorporate real schedulers and their constraints.

The best way to solve an application schedulability problem is analytical. With a full list of events, their deadlines and processing times, their periodicity or stochastic distribution, it is possible to model an ideal or a real real-time scheduler and to schedule the event handling at any given frequency. For example, rate-monotonic scheduling [5], deadline-monotonic scheduling [1] or their adaptation for DVS/DFSed system [4][6][11] could be used. However, the significant obstacle to applying this solution is that obtaining the full list of all events and all constraints in a typical mobile device is difficult. For example, wireless network related events, their deadlines and processing times can be obtained only by analyzing network protocols, processor communication with network interface hardware and so on. The increasing functionality and introduction of multiple real-time dependent interfaces, such as Bluetooth, WLAN, IrDA, GPS, and so on, in mobile devices make the task above very complicated. Furthermore, the limitations on real-time tasks assumed in literature such as periodicity or low-bound inter-arrival time intervals cannot always be assumed in real world systems. To summarize, analytical schedulability solution is good, but impossible to obtain in many real world cases.

When analytical solution is not available, the application schedulability is determined experimentally. The application is implemented and executed at different frequencies on hardware that allows such frequency variation. Alternatively, the application could be executed on a whole-system emulator that supports real-time simulation of any internal or external events and that allows frequency variation [10]. However, sometimes neither the available hardware nor the simulation environment supports the frequency changes. Sometimes the emulators support frequency changes but cannot model all needed real-time constraints. What can be done in such a case? For example, we wanted to determine the effects of processor frequency decrease in a mobile phone before the actual processor with frequency scaling was available. We had the software and the hardware available, but we could not change the processor frequency. This meant that we could not achieve our goals using any of the methods mentioned above. Consequently, we proposed, implemented, and used an approach that estimates the application's schedulability when only hardware with no frequency change capability is available. Our approach is described in the next section.

## 3 EXPERIMENTAL SCHEDULABILITY ESTIMATION

A simple estimate of schedulability is to determine the workload—the percentage of time during which a processor is active—at the maximum processor frequency and to expand this workload proportionally to the frequency reduction. This approach, for example, would indicate that a system with a 50% or lower workload at the maximum frequency is schedulable at 1/2 maximum frequency, since reducing the frequency in half raises the workload to 100%. Unfortunately, this approach is too simplistic for reliable schedulability determination. With the presence of real-time constraints that can be broken by a frequency reduction, the possible workload expansion is not a reliable indication of schedulability. We propose a more exact approach that takes care of checking the preservation of real-time constraints.

To estimate the schedulability of an application, we take the hardware device on which the application will be run and introduce additional software workload that approximates the processor frequency decrease on the device. We introduce a *slowdown* routine that runs at the highest priority in the shortest possible bursts. This routine has to run at the highest priority so that all software is affected by the slowdown. Also it has to run in the shortest possible bursts to minimize the distortion between the system behavior under slowdown routine and the system with actual frequency decrease. The slowdown routine can be implemented as an interrupt handler routine, since interrupt handlers usually have the highest priority in the system. Some operating systems contain executable elements with priorities higher than those of the interrupt handlers. In such operating systems, the slowdown routine has to have higher priority than these executable elements or these executable elements should be slowed down by other means, such as instrumentation.

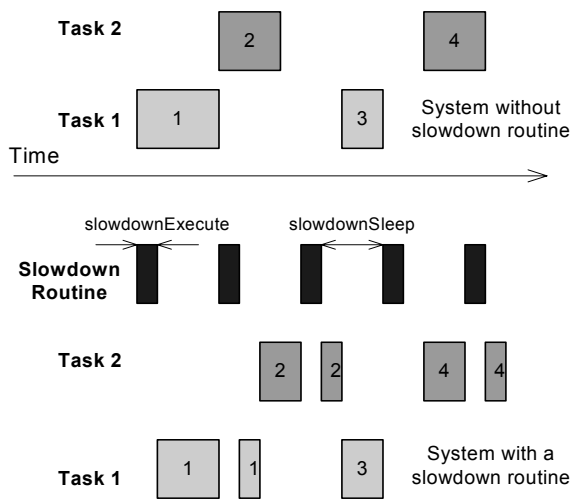

**Figure 1. Slowdown routine and task scheduling**

Figure 1 shows a system with a 25% slowdown routine added. The slowdown routine performs some work for slowdownExecute time and then exits with a timer set to awaken it after slowdownSleep time. The slowdownExecute and the slowdownSleep times should be set to the shortest possible intervals—possibly one clock tick depending on the operating system and its services—to have minimum distortion compared to the system with real frequency decrease. Compared to Figure 1, the slowdown routine would be active for very short times and very frequently, which is not illustrated in our simple example. To achieve less than 2% variation in overhead, the slowdown routine should be activated at least 10 times per task activation. In our system, the slowdown routine activated at least 10 times per original task activation for 64% of task invocations. The slowdownExecute and slowdownSleep can be set to different values to produce different overhead percentages corresponding to different simulated frequencies and different processor slowdowns. The slower processor approximation is inexact, because if some application task has real-time requirements that make this task to run during the time when the slowdown routine is active, such requirements will be violated even though they would not be necessarily violated in a system with the decreased frequency. For example, in Figure 1, the block 1 of task 1 is both moved in time and split into two because of the slowdown routine. It is possible that this block has a real-time constraint that is violated because it finishes 2 time measures later than in the original schedule. In a reduced frequency processor, such block may finish only 1 time measure—25%—later, which may still be acceptable. In another situation, if some task was active for less than slowdownSleep time, this task could be scheduled during the slowdownSleep interval and suffer no slowdown at all, though the same task would be slowed down if the processor frequency were reduced. In our system about 5% of task activations were shorter than slowdownSleep time and were not slowed down because of the reason above. Even with this observation, we consider that the method provided a reasonable schedulability estimate for our systems. However, other users of this method need to consider whether task activations in their system are much longer than slowdownSleep time and if not, whether the schedulability estimates provided by the method are still acceptable.

The difference between an approximation and an actual reduced-frequency processor diminishes, as the slowdownExecute and slowdownSleep times get smaller. The difference is smallest when slowdownExecute is equal to 1 processor clock tick. This necessitates keeping the slowdown routine work and idle times as short as possible to minimize the interference with other tasks. The method becomes less accurate when slowdownSleep is much larger than slowdownExecute or slowdownExecute is much larger than slowdownSleep, which is not usually the case for DVS/DFS frequency changes that fall into the interval of 10-90%. We executed our system with 10-90% slowdown.

Although short slowdown routine work time introduces a lot of context switching, we incorporate the memory, cache and other effects resulting from the context switches into the slowdownExecute/(slowdownSleep + slowdownExecute) ratio that indicates the frequency decrease. slowdownSleep/(slowdownSleep + slowdownExecute) multiplied by original frequency indicates the modeled processor frequency.

The approach above introduces uniform overhead over time, since slowdownExecute/(slowdownSleep + slowdownExecute) ratio remains constant during the application execution. However, our approach can be used also to determine the schedulability of the system under variable frequency. For example, one of the DVS/DFS control algorithms can be used to set the frequency per task or per time interval. This would be approximated by changing the slowdownExecute/(slowdownSleep + slowdownExecute) ratio of the slowdown routine for the task or the time interval to correspond to the frequency given by the DVS/DFS control algorithm. Variable frequency does not pose any additional difficulties for the estimation approach. The only condition necessary for its application is that the time between two processor frequency changes was much larger than slowdownExecute + slowdownSleep. Otherwise, the overhead introduced by the slowdown routine would not closely approximate the frequency decrease. Lee and

Krishna [5] noted that the DVS/DFS mode switching could be performed in microsecond range. Such DVS/DFS switch delay is insignificant and we do not consider it in the variable frequency schedulability estimation.

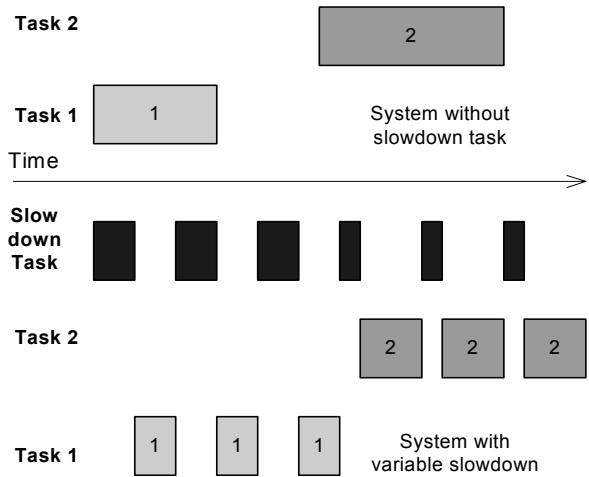

**Figure 2. Per task slowdown**

Figure 2 shows the possible scheduling with a variable per task slowdown. In it, task 1 is scheduled at 50% maximum frequency approximated by the 50% slowdown. During execution of task 2, the slowdown is only 25% simulating the execution at 75% of the maximum frequency. The figure is meant only as illustration and does not satisfy the condition above requiring the frequency change interval to be much larger than slowdownExecute + slowdownSleep. For example, in systems we analyzed, we used frequency change intervals 50-1000 times longer than slowdownExecute + slowdownSleep.

With our approach implemented, there is still a question how to determine if the resulting system is schedulable, i.e. if it functions correctly at the reduced frequency. The answer to this question can be found either in the formal system validation or, more often, in system testing with a comprehensive test suite. A complete test suite needs to be selected to obtain reliable test results and with them the schedulability estimate. Similar testing or validation needs to be done in other experimental schedulability estimation methods. The test suite can be executed at different frequencies using a binary partitioning to find the lowest frequency at which the test suite succeeds. In particular, testing could start at 50% slowdown and increase to 75% slowdown if the test suite succeeds or decrease to 25% slowdown if the test suite fails. Further binary partitioning could continue until the schedulable frequency is known as precisely as needed. Alternatively, the first test could start at the frequency corresponding to average workload, since such frequency should be close to the actual schedulable frequency. Such start could decrease the number of test runs. A more sophisticated approach is needed if the goal is to find an optimal DVS/DFS control algorithm that can change the frequency dynamically. This is left for the future work.

We have implemented the slowdown approach described above on several different mobile phones and have used it to determine the schedulability of various mobile phone applications including phone book browsing, games, web browsing, SMS message sends and receives and phone calls at various frequencies. The applications were schedulable at certain frequencies and became unschedulable at certain lower frequencies. The frequencies at which mobile phone applications are schedulable are not provided due to confidentiality reasons. We have also implemented the variable slowdown approach. In our implementation and system test executions, we observed less than 5% slowdown variation, i.e. the difference of the observed slowdown from the specified slowdown level during the execution of test programs. We are confident that our method provides an estimate for system schedulability at lower frequencies that could be used to evaluate DVS/DFS applicability.

## 4 ANALYSIS OF SCHEDULABILITY ESTIMATION APPROACH

Our approach is more exact than assuming applications to be schedulable at the frequency equal to average workload expressed as a fraction and multiplied by original frequency, since such number ignores any real-time constraints.

Although our approach can disturb the periodicity of lower priority tasks, because higher priority tasks will occupy more processor time, this effect also occurs in a genuine frequency decrease and is not an artifact of our approach. To function correctly at multiple processor frequencies and in our approach, tasks and operating system time services, such as timers, should refer to the absolute time and not to processor frequency dependent time values.

Our approach also improves upon the Weiser's et al. [16] and Govil's et al. [3] calculations of the DVS/DFS induced delays. Though the calculation of such delays provides a numerical estimate of schedulability, it does not take into account any real-time constraints that cannot tolerate delays. Weiser et al. [16] and Govil et al. [3] also argue that any workload cycles spilled over from one DVS/DFS interval to the next should be avoided. However, Pering et al. [10] and we argue that a system delay has to be avoided only if it breaks the real-time constraints or if it slows down the user interface. While Pering et al. [10] consider only such delays as waiting for an audio/video packet, UI event processing, our approach is more general since it considers the system schedulable only if it passes all testing criteria.

Weiser et al. [16] noticed that some task delays are "hard", i.e. sometimes a task needs to wait a set amount of time for a disk read or a network event (Figure 3, 1$^{st}$ diagram). Such gaps cannot be filled in a system with reduced frequency, as illustrated in the 2$^{nd}$ diagram in the Figure 3. DVS/DFS algorithms sometimes do not correctly deal with such hard delays, expecting as in the 3$^{rd}$ diagram in Figure 3 to fill the gap with computation.

Our approach correctly deals with the hard delays, since it slows down the system independently of the hard delay existence. By doing this it does not fill the hard delays. If our approach is used with a DVS/DFS control algorithm, the algorithm itself has to deal with hard delays correctly by not decreasing the frequency to fill the hard waiting gaps. If the DVS/DFS algorithm incorrectly tries to fill hard delays, such errors will be detected during the testing.

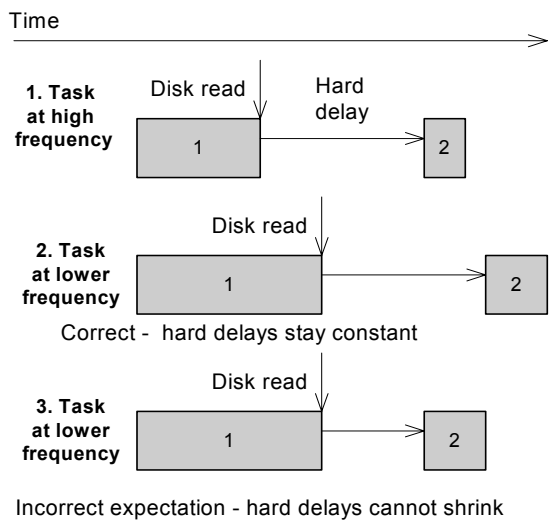

**Figure 3. Hard delays**

Our approach is weaker than a formal schedulability analysis. If researchers can obtain and formally analyze the real-time requirements and constraints of a system, they can create an ideal schedule that conforms to these requirements. Such a schedule would indicate the ideal schedulability. Unfortunately, most of the time such analysis is extremely complicated.

## 5 RELATED WORK

DVS/DFS was proposed by Weiser et al. [16]. They also proposed the concept of "soft" and "hard" delays. Govil et al. [3] elaborated on Weiser et al., proposed and tested a variety of DVS/DFS control algorithms on Weiser's traces. Pering et al. [10], Sinha and Chandrakasan [15] suggested additional DVS/DFS control algorithms. Pouwelse et al. [12][13][14] suggested DVS/DFS control based on information from power aware applications.

Our work is not applicable in the situations where DVS/DFS was already implemented on real systems, since in these cases the schedulability can be determined on the implemented platform and does not need to be estimated. We target systems with hardware that does not yet support frequency change.

Both Weiser et al. and Govil et al. use fine-grained delay measures. They assume that any workload cycles spilled over from one DVS/DFS interval to the next contribute to the delay. Pering et al. [10] paper expands on the work by Weiser et al. and Govil et al. They use implementable variations of DVS/DFS algorithms in a simulation environment for mobile device data suite: address book browsing, real-time audio and MPEG decompression. They introduce a higher-level delay metric that does not penalize processing during the time when system waits for an audio/video packet. They also do not penalize the processing delays unnoticeable in the user interface. Our approach uses a similar metric, since it considers the system schedulable if none of the testing criteria—including usability criteria—of the system are broken.

Lee and Krishna [5] and Gruian [4] propose the DVS/DFS algorithms that ensure schedulability for a known set of real-time tasks with specified periods, worst-case execution times and deadlines. Our work assumes a system where some of these parameters are unknown.

Lorch and Smith [7] propose an optimization applicable to any DVS/DFS algorithm based on work distribution in the executed task. Their work is related to ours in the sense that they propose an addition to any DVS/DFS algorithm. Similarly our schedulability estimation can be performed under any DVS/DFS algorithm.

Martin and Siewiorek [8] noted some non-linear battery and memory effects that complicate the energy savings due to voltage and frequency reduction. Since our approach does not estimate energy savings, only schedulability, we ignore Martin and Siewiorek's findings.

## 6 LESSONS AND FUTURE WORK

The important lesson from estimating application schedulability is that application design strongly influences the schedulability of the system. To generalize this, we claim that improving the performance of an application also improves its other characteristics: energy consumption—by leaving more time for a processor to sleep, schedulability—by leaving more unconsumed processor time and executing tighter than any of the real-time constraints, and so on. The general optimization rule of thumb that there is no point in optimizing a system if the user does not see a difference in the user interface is shown to be wrong. Even if the user does not see a difference, the energy consumption and schedulability can be further improved by improving performance, which will affect users through the energy conservation. This needs to be further studied and presented to the system designers perhaps as design patterns [2] for schedulability improvement.

Another area for further study is real-time constraints and acceptable task latency deadlines, since they also affect schedulability and are not necessarily improved by the performance improvements. For example, if the real-time constraints can be relaxed without changing the performance, the system's schedulability will improve.

## 7 CONCLUSIONS

This paper presents a schedulability problem that is the basis for DVS/DFS applicability. We describe a new approach to the experimental estimation of application and system schedulability. The basis of the approach is the introduction of additional workload in an artificial highest priority routine to approximate the processor frequency decrease. Thus we applied the idea of using extra workload in the new context of simulating frequency decrease similar to actual DVS/DFS. We have implemented this approach and analyzed mobile phone application schedulability in the situation where no DVS/DFS capability was available and no other analysis methods could be used. The paper presents our approach, its analysis, lessons learned and the future work. We believe that the new schedulability approach will be helpful for DVS/DFS analysis and introducing DVS/DFS into various systems and applications.

## 8 ACKNOWLEDGEMENTS

We thank the people from Nokia Mobile Phones who supported this research. We thank Karel Driesen and anonymous reviewers for valuable comments on this paper.

## 9 REFERENCES

[1] N.C. Audsley, A. Burns, M.F. Richardson, A.J. Wellings, Hard Real-Time Scheduling: The Deadline-Monotonic Approach, *Eighth IEEE Workshop on Real-Time Operating Systems and Software*, pp. 133-137, 1991.

[2] E. Gamma, R. Helm, R. Johnson, J. Vlissides, *Design Patterns Elements of Reusable Object-Oriented Software*. Addison-Wesley, Reading, Massachusetts, 1994.

[3] K. Govil, E. Chan, H. Wasserman, Comparing Algorithms for Dynamic Speed-Setting of a Low-Power CPU, *Proceedings of the First Annual International Conference on Mobile Computing and Networking, ACM Press*, pp.13-25, 1995.

[4] F. Gruian, Hard Real-Time Scheduling for Low-Energy Using Stochastic Data and DVS Processors, *Proceedings of the International Symposium on Low Power Electronics and Design 2001*, Huntington Beach (CA), US, pp. 46-51, August 6-7, 2001.

[5] Y.-H. Lee, C.M. Krishna, Voltage-clock Scaling for Low Energy Consumption in Real-time Embedded Systems, *Proceedings of the Sixth International Conference on Real-Time Computing Systems and Applications*, pp 272-279, Hong Kong, China, December 1999.

[6] C. L. Liu, J. W. Layland, "Scheduling algorithms for multiprogramming in a hard-real-time environment", *Journal of the ACM*, vol. 20, no 1, pp. 46-61, 1973.

[7] J. Lorch, A.J. Smith, Improving dynamic voltage scaling algorithms with PACE. *Proceedings of the ACM SIGMETRICS 2001 Conference*, Cambridge, MA, pp. 50–61, June 2001.

[8] T.L. Martin, D.P. Siewiorek, The Impact of Battery Capacity and Memory Bandwidth on CPU Speed-Setting: A Case Study, *Proceedings of the International Symposium on Low Power Electronics and Design 1999*, San Diego, USA, pp. 200-205, 1999.

[9] Palm Inc., www.palm.com, 2002.

[10] T. Pering, T. Burd, R. Brodersen, The simulation and evaluation of dynamic voltage scaling algorithms, *Proceedings of the International Symposium on Low Power Electronics and Design 1998,* pp. 76-81, August 1998.

[11] P. Pillai, K.G. Shin, Real-Time Dynamic Voltage Scaling for Low-Power Embedded Operating Systems, *Proceedings of 18th ACM Symposium on Operating Systems Principles (SOSP'01)*, pp. 89-102, Banff, Alberta, Canada, October, 2001.

[12] J. Pouwelse, K. Langendoen, H. Sips, Dynamic Voltage Scaling on a Low-Power Microprocessor, *Proceedings of the 7th International Conference on Mobile Computing and Networking (Mobicom)*, pp. 251-259, Rome, Italy, July 2001.

[13] J. Pouwelse, K. Langendoen, H. Sips, Application-directed voltage scaling, *IEEE Transactions on Very Large Scale integration (TVLSI)*, September 2002.

[14] J. Pouwelse, K. Langendoen, H. Sips, Energy priority scheduling for variable voltage processors, *Proceedings of the International Symposium on Low Power Electronics and Design 2001*, Huntington Beach (CA), US, pp. 28-33, August 6-7, 2001.

[15] A. Sinha, A. Chandrakasan; "Dynamic Voltage Scheduling Using Adaptive Filtering of Workload Traces", *Proceedings of the 14th International Conference on VLSI Design*, Bangalore, India, January 2001.

[16] M. Weiser, B. Welch, A. Demers, S. Shenker, "Scheduling for Reduced CPU Energy," *Proceedings of the 1st USENIX Symposium on Operating Systems Design and Implementation*, pp. 13-23, November 1994.